\UseRawInputEncoding
\documentclass[10pt,A4paper,amsmath,amssymb,aps,
etoolbox,floatfix,
longbibliography,nofootinbib,
noshowkeys,noshowpacs, 
onecolumn,
preprintnumbers,prl,reprint,
showpacs,
]
{revtex4}

\newcommand{\bea}{\begin{eqnarray}}
\newcommand{\beq}{\begin{equation}}

\newcommand{\eea}{\end{eqnarray}}
\newcommand{\eeq}{\end{equation}}

\usepackage{amsmath}
\usepackage{amsfonts}
\usepackage{amssymb}
\usepackage{bm}       
\usepackage{caption}
\usepackage{color}
\usepackage{dcolumn}  
\usepackage{epsfig}
\usepackage{float}
\usepackage{graphicx} 
\usepackage{hyperref}

\begin{document}

\title{The Heisenberg limit at cosmological scales}

\author{Alessandro D.A.M. Spallicci,\textsuperscript{a,b,c,d}\footnote{spallicci@cnrs-orleans.fr 
}, 
Micol Benetti,\textsuperscript{e,f,g}\footnote{micol.benetti@unina.it},
Salvatore Capozziello\textsuperscript{e,f,g,h}\footnote{salvatore.capozziello@unina.it}
}

\affiliation{
\mbox{\textsuperscript{a}Institut Denis Poisson (IDP) UMR 7013}\\
\mbox{Universit\'e d'Orl\'eans (UO) et Universit\'e de Tours (UT)}\\
\mbox{Centre National de la Recherche Scientifique (CNRS)}\\
\mbox{Parc de Grandmont, 37200 Tours, France}
\vskip2pt
\mbox{\textsuperscript{b}Laboratoire de Physique et Chimie de l'Environnement et de l'Espace (LPC2E) UMR 7328}\\
\mbox{Centre National de la Recherche Scientifique (CNRS)}\\
\mbox{Universit\'e d'Orl\'eans (UO)}\\
\mbox{Centre National d'\'Etudes Spatiales (CNES)}\\
\mbox {3A Avenue de la Recherche Scientifique, 45071 Orl\'eans, France}
\vskip2pt
\mbox{\textsuperscript{c}UFR Sciences et Techniques}\\
\mbox{Universit\'e d'Orl\'eans (UO)}\\
\mbox{Rue de Chartres, 45100 Orl\'{e}ans, France}
\vskip2pt
\mbox{\textsuperscript{d}Observatoire des Sciences de l'Univers en region Centre (OSUC) UMS 3116}\\
\mbox{Universit\'e d'Orl\'eans (UO)}\\
\mbox{1A rue de la F\'{e}rollerie, 45071 Orl\'{e}ans, France}
\vskip2pt
\mbox{\textsuperscript{e}Dipartimento di Fisica Ettore Pancini, Universit\`a degli Studi di Napoli, Federico II (UNINA)}\\
\mbox{Complesso Universitario Monte S. Angelo, Via Cinthia 9 Edificio G, 80126 Napoli, Italy}
\vskip2pt
\mbox{\textsuperscript{f}Istituto Nazionale di Fisica Nucleare (INFN), Sezione di Napoli}\\
\mbox{Complesso Universitario Monte S. Angelo, Via Cinthia 9 Edificio G, 80126 Napoli, Italy} 
\vskip2pt
\mbox{\textsuperscript{g}Scuola Superiore Meridionale}\\
{\mbox Largo San Marcellino 10, 80138 Napoli, Italy} 
\vskip2pt
\mbox{\textsuperscript{h}Laboratory  for  Theoretical Cosmology, Tomsk State  University  of Control  Systems  and  Radioelectronics (TUSUR)}
\mbox{40 prospect Lenina, 634050 Tomsk, Russia.}
}

\begin{abstract}
{For} an observation time {equal to} the universe age, the Heisenberg principle  fixes the value of the smallest measurable mass at $m_{\rm H}=1.35 \times 10^{-69}$ kg and prevents to probe the masslessness for any particle {using a balance}. The corresponding reduced Compton length to $m_{\rm H}$ is $\lambdabar_{\rm H}$, and 
represents the length limit beyond which masslessness cannot be proved using a metre ruler. In turns, $\lambdabar_{\rm H}$ is equated to the luminosity distance $d_{\rm H}$ which corresponds to a red shift $z_{\rm H}$. When using the Concordance-Model parameters, we get $d_{\rm H} = 8.4$ Gpc and  $z_{\rm H}=1.3$. Remarkably, $d_{\rm H}$ falls quite short to the radius of the {\it observable} universe. According to this result, tensions in cosmological parameters could be nothing else but due to comparing data inside and beyond 
$z_{\rm H}$.  Finally, in terms of quantum quantities, the expansion constant $H_0$ reveals to be one order of magnitude above the smallest measurable energy, divided by the Planck constant.\end{abstract}

\date{11 December 2021}


\maketitle

The Heisenberg principle, considered at cosmological scales, has been adopted \cite{capozziello-benetti-spallicci-2020} to address the so-called Hubble tension 
\cite{divalentino-etal-2021a,divalentino-etal-2021b,gurzadyan-stepanian-2021,moussa-shababi-ramahan-kumardey-2021,perivolaropoulos-skara-2021, yang-etal-2021,vagnozzi-2021,aghababaei-moradpour-vagenas-2021,petronikolou-basilakos-saridakis-2021,riess-etal-2021} taking into account  the differences between kinematic and dynamical measurements of the {Hubble-Lema\^itre} constant\footnote{The International Astronomical Union has adopted the terminology Hubble-Lema\^itre constant \cite{IAU-2018}. We recall the contributions by Humason \cite{humason-1929,humason-1931,hubble-humason-1931}.}
$H_0$. Here, we want to investigate if  the Heisenberg principle has  other roles at  cosmological scales and if there are implications for fundamental physics. In other words, we intend to broadly explore the issue of measurability and the relations between quantum mechanics and cosmology. Are measurements always possible in an expanding universe? Does the Heisenberg principle prevent the possibility of measurements at some given cosmological scale for those quantities related to the principle? Since most of cosmological observations are performed with light, awareness of the implications of photon properties is essential.

The Compton length 
represents the distance at which the concept of a single point-like particle breaks down since the particle cannot be localised within this distance. In the reduced form, it is given by

\beq
\lambdabar_{\rm C} = \frac{\hbar}{m c}~. 
\label{comptonlength}
\eeq
where $\hbar$ is the reduced Planck constant, $m$ the mass of the particle and $c$ the speed of light. 
Thus, the Compton length is the quantum counterpart of mass. It means that defining a Compton length means defining masses of particles as well as any effective mass.
When we deal with Bosons, the Compton length assumes the further meaning of scale of the interaction range. 
Gravitation and electromagnetism are interactions without a Yukawa term \cite{scharffgoldhaber-nieto-2010}. As long as photons and gravitons are massless, $\lambdabar_{C}$ remains infinite. However, a tiny mass would reduce the interaction at a distance $r$ by a factor $e^{- r/\lambdabar_{C}}$. For photons, this observation was made by Schr\"odinger \cite{schrodinger-1943}. 

The group velocity dispersion and the deviations from the Gauss law on the electric field divergence or from the Amp\`ere-Maxwell law on the magnetic field curl, caused by a photon mass and thereby to a non-infinite Compton length, were first written by de Broglie \cite{debroglie-1922,debroglie-1936}.

Tinier and tinier photon mass upper limits may be sought through testing at larger and larger distances \cite{scharffgoldhaber-nieto-2010}. 
Precision local experiments on mass might be replaced by exploring the possible finitude of the range of the interaction. Indeed, experimentally, a demanding precision measurement on a Boson mass may be circumvented if looking for deviations of {the} potential from the $1/r$ inverse law.

Turning to the Heisenberg indeterminacy or uncertainty principle\footnote{Herein, we use indeterminacy when we address massiveness as opposed to masslessness and uncertainty when we refer to measurement accuracy. Principle is used when referring to the general concept rather than an estimate.} 

\begin{align}
\Delta p \Delta x = \Delta E \Delta t > {\hbar}/{2}~,
\label{heisenberg}
\end{align}
as $ E = m c^2$, a lower limit on the measurability of the rest mass of any particle is implied. 
The Heisenberg principle in the energy-time formulation already appeared in his earliest work \cite{heisenberg-1927}, and it states that 
there is a smallest measurable quantity of mass $m_{\rm H}$ (the letter H as lower index stands for Heisenberg). Today estimate is $1.35 \times 10^{-69}$ kg  for an observation time equal to the universe age\footnote{Photons were emitted at decoupling about 370.000 years after the Big Bang; photons by galaxies some hundred million years later.} of about $13.8$ Gy. 

The principle implies an uncertainty in measuring massive particles but also the intrinsic indeterminacy or impossibility of asserting {\it masslessness}. The same principle indicates the minimum time necessary to achieve a given precision. Nevertheless, tests might be carried out observing a classical effect by an ensemble of Bosons, thereby partly circumventing the constraints of a limited observation time. For discussions on the Heisenberg principle in the energy-time form, see \cite{ballentine-1970,bush-1990a,bush-1990b,kobe-aguileranavarro-1994,bush-etal-2007,bush-2008}.   


In \cite{capozziello-benetti-spallicci-2020},  we portrayed the  Compton length as the concept relating the luminosity distance with the Heisenberg principle. Indeed, inserting $m_{\rm H}$ in Eq. (\ref{comptonlength}), the reduced Compton length $\lambdabar_{\rm C}$ becomes $\lambdabar_{\rm H}$ and it assumes the meaning of mass limit expressed in metres. In other words, the photon {\it masslessness} cannot be proved also through the corresponding Compton length. As there is not a balance measuring below $m_{\rm _H}$, there is not a test,  at a given distance $r$, that can measure an attenuation factor smaller than $e^{-r/\lambdabar_{\rm H}}$ for the electromagnetic interaction.   

Before dealing with cosmology, we remark that testing of massive electromagnetism\footnote{De Broglie estimated the photon mass below $10^{-53}$ kg through group velocity dispersion 
\cite{debroglie-1923,debroglie-1924}. 
After one century of painstaking experiments, the Particle Data Group (PDG) \cite{zyla-etal-2020} sets the upper limit just one order of magnitude below, at $m < 10^{-18}$~eV c$^{-2}$ or 
$1.8 \times 10^{-54}$~kg in the solar wind, but see \cite{retino-spallicci-vaivads-2016} for a critical assessment. Recently, limits up to  $5.1\times 10^{-51}$ kg were established estimating dispersion in Fast Radio Bursts (FRBs) \cite{xing-etal-2019}, more stringent than other attempts \cite{boelmasasgsp2016,boelmasasgsp2017}; for a review, see \cite{wei-wu-2021}. The best laboratory test - based on Coulomb's law - showed an upper limit of $1.6 \times 10^{-50}$ kg \cite{wifahi71}, while a space mission was conceived targeting dispersion in the very-low radio frequencies \cite{bebosp2017}.} and massive gravitation\footnote{Massive gravity was first proposed by Fierz and Pauli \cite{fierz-pauli-1939}, and later in different contexts including alternatives to dark matter \cite{capozziello-basini-delaurentis-2011}. For gravitons, the PDG \cite{zyla-etal-2020} sets the upper limit at $6 \times 10^{-32}$ eV or $1.1 \times 10^{-67}$ kg, obtained from gravitational lensing \cite{choudury-etal-2004}. } 
has led to significant mass upper limits \cite{scharffgoldhaber-nieto-2010}. 

Photons are the only massless particles in the Standard-Model (SM), but acquire an effective mass in the SM Extension (SME) 
\cite{bonetti-dossantosfilho-helayelneto-spallicci-2017,bonetti-dossantosfilho-helayelneto-spallicci-2018,helayelneto-spallicci-2019,spallicci-etal-2021}. For massive photons, gauge invariance, renormalisability and charge conservation may hold \cite{capozziello-benetti-spallicci-2020}.   

{We now turn to cosmology and recall that $H_0$ is}  valued $74.03 \pm 1.42$ km/s per Mpc \cite{riess-etal-2019} by current {Supernova} measurements. On the other hand, such a constant is constrained {at} $66.88 \pm 0.92$ km/s per Mpc \cite{aghanimetal2020} using the Cosmic Microwave Background and the Baryonic Acoustic Oscillations data. The difference between the two values, named {\it{Hubble tension}}, is more than $4\sigma$ (actually $5\sigma$ \cite{riess-etal-2021}) and can be compared to the Heisenberg uncertainty \cite{capozziello-benetti-spallicci-2020}. Herein, for our purposes, we take an average value of $H_0=70$ km/s per Mpc. 

{The $m_{\rm H}$ value corresponds} to a reduced Compton length $\lambdabar_{\rm H}$  of 
$2.6 \times 10^{26}$ m that is 8.4 Gpc, Eq (\ref{comptonlength}). Incidentally, the today official upper limit on photon mass corresponds to a reduced Compton length of $1.9 \times 10^{11}$ m \cite{zyla-etal-2020}, somewhere between Jupiter and Saturn, thanks to the Voyager and Pioneer missions. 

We stated that the Compton length for a Boson is a measure of the interaction range. The latter appears best corresponded by the distance crossed by the photon from the source at emission time to the Earth at reception time. In an expanding universe such distance may be best corresponded by the luminosity distance, though such distance is defined through bolometric quantities.

\begin{figure}[H]
\begin{center}
\includegraphics[width=9cm]{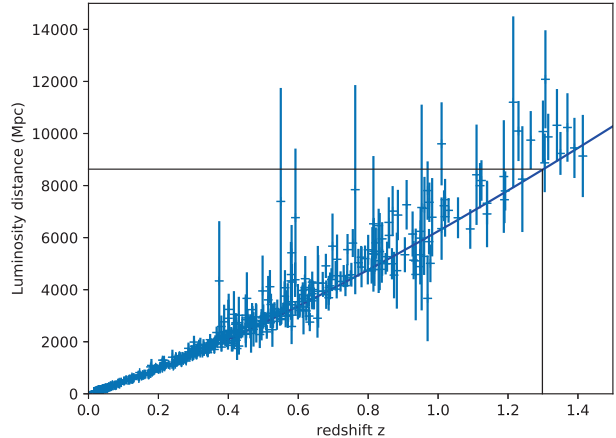}\\
\caption{Luminosity distance (Mpc) as function of $z$ plotted against SN Union 2.1 data \cite{amanullah-etal-2010}. For the (2,1) Pad\'e approximant, Eq. (\ref{dLpade21}), using the Concordance-Model parameters (see footnote 7): $H_0=70$ km/s per Mpc, $\Omega_{\rm m} = 0.3$ and $\Omega_{\Lambda} = 0.7$, we get $q_0 = - 0.29, j_0 = 0.55$ \cite{capozziello-et-al-2019,capozziello-et-al-2020}. 
The horizontal line represents the reduced Compton length $\lambdabar_{\rm H} = 8.4$ Gpc, corresponding to the smallest measurable mass, $1.35 \times 10^{-69}$ kg. The vertical line at $z_{\rm H} = 1.35$, represents the Heisenberg limit. }
\label{fig1}
\end{center}
\end{figure}

We aim to identify the red shift $z_{\rm H}$ corresponding to $\lambdabar_{\rm H}$, without relying on a specific cosmological model. We  thereby compute the luminosity distance through a cosmographic {parametrisation}. Specifically, we can use
Pad\'e approximants as they have proven to be efficient in
{parametrising} cosmic distances whilst reducing error propagation
for $z > 1$  \cite{capozziello-et-al-2019,capozziello-et-al-2020}. The idea consists in expressing the luminosity distance as a ratio of Taylor series and probe its convergence against data. In particular, the luminosity distance computed out of the (2,1) Pad\'e
 approximant \footnote{The Taylor expansion of a generic function $f(z)$ is 
\beq
f(z)=\sum_{i=0}^\infty c_i z^i~,
\eeq where $c_i=f^{(i)}(0)/i!$, whereas the $(n,m)$ Pad\'e approximant of $f(z)$ is defined as rational polynomial by
\beq
P_{n,m}(z)=\dfrac{\displaystyle{\sum_{i=0}^{n}a_i z^i}}{1+\displaystyle{\sum_{j=1}^{m}b_j z^j}}\,.
\label{eq:def Pade}
\eeq} 
reads
\beq
d_{\rm {L}} = \frac{c}{H_0} \frac{6(q_0 - 1) - z[5 + 2j_0 - q_0(8 + 3q_0)]}{2q_0(3 + z + 3q_0z) -2(3 + z + j_0z)}\, z =   \frac{c}{H_0} P_{2,1} \, z~,
\label{dLpade21}
\eeq
where $q_0$ and $j_0 $ are the deceleration and jerk parameters, respectively. As shown in \cite{capozziello-et-al-2020}, this approximant well fits data at high and low red shifts. 
The (2,1) Pad\'e approximant can be checked by  Supernova (SN) data from the Union 2.1  catalogue  \cite{amanullah-etal-2010}, Fig. \ref{fig1}.

The check being successful\footnote{The check consists in fitting observational data with the Pad\'e approximants. The Pad\'e-approximant based cosmography reproduces observed distances with analytical models. The latter include a set of cosmological parameters which determine the coefficients of the approximants. In Fig. \ref{fig1}, we have used Concordance-Model parameters ($\Omega_{\rm m} = 0.3$ and $\Omega_{\Lambda} = 0.7$),  and thereby we have identified the coefficients $q_0$ and $j_0$. 
We could have departed from the Concordance-Model parameters and then obtained other values of the Heisenberg radius, but without relevant consequences for our work.}, we deduce that the $z$, corresponding  to $\lambdabar_{\rm H} = 8.4$ Gpc is 1.35, Fig. \ref{fig1}, for the Concordance-Model parameters. Our aim is now to interpret such a result.  

The distance of 8.4 Gpc falls into the radius of the observable universe\footnote{The radius of the observable universe is usually expressed as comoving distance, 14.3 Gpc, roughly at $z=1100$. For a flat universe, 
$\Omega_{\rm k} = 0$, the luminosity distance is $(1+z)$ times the comoving distance, achieving thereby extremely large values for increasing $z$.}, that is the cosmological region causally connected by light. We can now reinterpret $\lambdabar_{\rm H}$ as the {\it Heisenberg radius}, the cosmological length beyond which the indeterminacy principle prevails\footnote{McCrea \cite{mccrea-1960a} used the term uncertainty for pointing out the consequences of the finitude of the speed of light on the observability of the universe and did not address the Heisenberg principle in any manner. {In \cite{albrecht-phillips-2014,sahlen-2017,beltrame-2021}, the origin of probability in cosmology, the relations to inflation, the boundaries of knowledge are addressed. The latter issue is closest to our work. Further discussions on the limits of knowledge, possibly in connection with G\"odel incompleteness theorem, go well beyond the aims of this paper.}}.

\begin{figure}[H]
\begin{center}
\includegraphics[width=8cm]{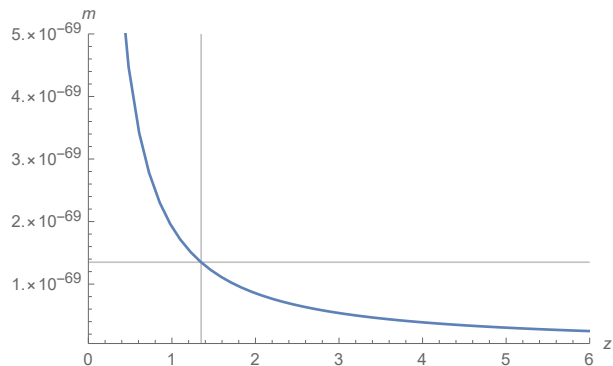}
\caption{Measurable mass (kg) as function of $z$, Eq. (\ref{mpade21}). For the (2,1) Pad\'e approximant, Eq. (\ref{dLpade21}), using the Concordance-Model parameters (see footnote 7): $H_0=70$ km/s per Mpc, $\Omega_{\rm m} = 0.3$ and $\Omega_{\Lambda} = 0.7$, we get $q_0 = - 0.29, j_0 = 0.55$ \cite{capozziello-et-al-2019,capozziello-et-al-2020}. 
The horizontal line represents the smallest measurable mass, $1.35 \times 10^{-69}$ kg. The vertical line at $z_{\rm H} = 1.35$, represents the Heisenberg limit.}
\label{fig2}
\end{center}
\end{figure}

Combining Eqs. (\ref{comptonlength})  and (\ref{dLpade21}), the measurable mass as function of $z$ is given by, Fig. \ref{fig2}
\beq
\left. m = \frac{ \hbar}{c  \lambdabar_{\rm C}}\right|_{\lambdabar_{\rm C} = d_{\rm L}}
\label{mpade21}
\eeq 

Equation (\ref{mpade21}) shows a peculiar situation, mentioned before: any measurement beyond $\lambdabar_{\rm H}$ will not lower the mass limit $m_{\rm H}$. This occurs well within the observable universe.

Perhaps, not surprisingly at this stage, we can relate the expansion constant $H_0$ to quantum quantities. The $H_0$ averaged and indicative value of 70 km/s per Mpc corresponds to $2.3 \times 10^{-18}$ m/s per metre. Such value may be derived from the Heisenberg  principle, Eq. (\ref{heisenberg}), in the energy-time form, recalling that the $H_0$ constant has the dimension of time$^{-1}$. Indeed 

\beq
H_0 = \frac{1}{\Delta t}=\frac{2 \Delta E}{\hbar} = \frac{2 m_{\rm H}c^2}{\hbar} = \frac{4 \pi m_{\rm H}c^2}{h} = 2.3 \times 10^{-18} {\rm m/s~per~metre}~.
\label{H0}
\eeq 
 
Equation (\ref {H0}) may suggest that the Hubble tension \cite{capozziello-benetti-spallicci-2020} is not accidental, but simply one facet of a possibly emerging evidence: the observation of the expansion reveals being a quantum measurement. Notably, $H_0$ reveals to be just one order of magnitude ($4\pi$) above the smallest measurable energy, divided the Planck constant $h$: 10\% of $H_0$ is at the same time the Heisenberg limit and the magnitude of the Hubble tension. 

%
%
%

According to our analysis, we can draw the following conclusions.
\begin{enumerate}
\item{Any measurement on photon mass through its reduced Compton length beyond $\lambdabar_{\rm H}$ will provide {\it at best} the Heisenberg limit, $1.35 \times 10^{-69}$ kg,}  
\item{Masslessness cannot be proved due to the Heisenberg principle, anywhere, either with a balance or with a metre ruler at any scale length. If there is no evidence of a particle being massive at a distance smaller than $\lambdabar_{\rm H}$, there is no way to ascertain its masslessness beyond, through its reduced Compton length.}  
\item{A basic question ensues: can we interpret light coming from cosmological sources as composed by massless photons, when we cannot prove their masslessness? The answer lies on whether computations and observations are consistent with such an assumption or not. This is applicable before and after $\lambdabar_{\rm H}$.}
\item {
One of the tenets of multi-messenger astronomy is comparing gravitational and electromagnetic signals, assuming that the latter travel at the speed of light $c$. This is done in view of discriminating gravitational waves from general relativity (speed $=c$), from gravitational waves of other gravitation theories (speed $\neq c$). 
Multi-messenger astronomy might need to take into account that gravitons and photons cannot be proved massless and thereby their speed is not $c$. Incidentally, at 
$\lambdabar_{\rm H}$, the delay of a massive photon, supposedly $10^{-54}$ kg - {actual} upper limit \cite{zyla-etal-2020} - with respect to a massless particle, reaches 10 ns at 1 GHz, Fast Radio Burst typical frequency \cite{boelmasasgsp2016,boelmasasgsp2017,bebosp2017}.}
\item{Assuming the universe as a standard physical (quantum) system, cosmological measurements before or after $\lambdabar_{\rm H}$, must be read with care. Issues as the $H_0$ tension point out that measurements inside and beyond $\lambdabar_{\rm H}$ could be not consistent. 
{Searching in the current state of observations an unequivocal trend between multiple different measurements at low red shift and multiple different measurements at high red shift appears the following step. Being the state of the art of measuring photon mass quite coarser than the Heisenberg uncertainty, rendering the items 1-2 beyond reach, our interpretation of the Hubble tension  \cite{capozziello-benetti-spallicci-2020} stands as the only indication of this trend, we are actually aware of.}}
\item{The Hubble-Lema\^itre constant $H_0$ is related to the Heisenberg principle and thereby determined by quantum quantities. }
\end{enumerate}

{\bf Acknowledgements} \\
ADAMS acknowledges the Erasmus+ programme for visiting the Universit\`a di Napoli, SC the {Universit\'e d'}Orl\'eans and Campus France for the hospitality. SC and MB acknowledge the Istituto Nazionale di Fisica Nucleare (INFN), sezione di Napoli, {\it iniziative specifiche} MOONLIGHT2 and  QGSKY. The authors thank O. Luongo (Frascati) for discussions. Finally, the authors are indebted to the referee for the detailed comments.

\bibliography{references_spallicci_211211}
\end{document}